\documentstyle[preprint,tighten,pra,aps]{revtex}
\begin{document}
\tightenlines
\draft

\def\bce{\begin{center}}
\def\ece{\end{center}}
\def\beq{\begin{eqnarray}}
\def\eeq{\end{eqnarray}}
\def\ben{\begin{enumerate}}
\def\een{\end{enumerate}}
\def\ul{\underline}
\def\ni{\noindent}
\def\nn{\nonumber}
\def\bs{\bigskip}
\def\ms{\medskip}
\def\tr{\mbox{tr}}
\def\wt{\widetilde}
\def\wh{\widehat}
\def\brr{\begin{array}}
\def\err{\end{array}}
\def\dsp{\displaystyle}

\title{Fluctuations of Quantum Fields \\ 
via Zeta Function Regularization}

\author{Guido Cognola$^{1,}$\footnote{email: cognola@science.unitn.it,
cognola@tn.infn.it},
Emilio Elizalde$^{2,}$\footnote{email:elizalde@ieec.fcr.es, 
elizalde@math.mit.edu} and
Sergio Zerbini$^{1,}$\footnote{email: zerbini@science.unitn.it,
zerbini@tn.infn.it}}

\address{$^1$ Dipartimento di Fisica, Universit\`a di Trento 
   \\ and Istituto Nazionale di Fisica Nucleare 
   \\ Gruppo Collegato di Trento, Italia}

\address{$^2$ Instituto de Ciencias del Espacio, IEEC/CSIC, 
IFAE,\\ and University of Barcelona, Spain}

\maketitle

\begin{abstract}

Explicit expressions for the expectation values and the  variances of some 
observables, which are bilinear quantities in the quantum fields on a
$D$-dimensional manifold,  
are derived making use of zeta function regularization.
It is found that the variance, related to the second functional variation 
of the effective action, requires a further regularization and 
that the relative  regularized 
variance turns out to be  $\frac{2}{N}$, where $N$ is the number of the 
fields, thus being independent on the dimension $D$.
Some illustrating examples are worked through.
The issue of the stress tensor is also briefly addressed.

\end{abstract}

\pacs{05.30.-d,05.30.Jp,11.10.Wx,11.15.Ex}


\section{Introduction}
\label{Form}

Vacuum fluctuations play an important role in many physical processes.
The Casimir effect is one of its most 
interesting physical manifestations and it has been experimentally verified. 
It is also well known that the
Casimir effect is related to the presence of a non vanishing vacuum energy 
(see for example \cite{birrell,moste}). This fact mainly occurs 
when one is 
dealing with non 
trivial space-times, 
where the spatial sections are topologically non trivial spaces or
manifolds with  boundaries.   

Another interesting issue where quantum vacuum effects are present is the 
Quantum Field Theory in curved space-time
\cite{birrell,fulling,wald}.
Recall that within the semi-classic approach to quantum gravity, the basic 
equation reads
\begin{eqnarray} 
G_{\mu \nu}+g_{\mu \nu}\Lambda = 8\pi G<T_{\mu \nu} >\:,
\end{eqnarray} 
where $G_{\mu \nu}$ is the Einstein tensor, $\Lambda$ the cosmological 
constant  and $T_{\mu \nu}$ is the vacuum 
expectation value of the matter stress tensor (we use $\hbar=c=1$). 
As a consequence,  
fluctuations of the stress tensor can induce fluctuations of the classical 
gravitational field and, in order to justify the semiclassical approximation, 
 it appears  very  important to  have reliable {\it a priori}
estimates of these fluctuations. 

Fluctuations of the stress tensor were studied in \cite{ford,ford1} 
making use of canonical 
methods. Fluctuations of Casimir forces were investigated in 
\cite{barton,eberlein}. Alternatively, other authors \cite{hu} have 
investigated 
the same problem by 
making use of zeta function regularization.  

With regard to this issue, it is well known that the 
notion of effective action 
(or effective potential) plays an important role as a powerful tool in
relativistic quantum field theory. This quantity, however,  is ill 
defined, since, within the Euclidean formulation and in the  
one-loop approximation, 
the one-loop effective action contains functional determinants of elliptic
operators, which have to be regularized. 
Zeta-function
regularization \cite{ray71-7-145,dowk76-13-3224,hawk77-55-133}, 
(see also \cite{kont95b} for the generalization to elliptic 
pseudo-differential 
operators and \cite{eliz94b,eli2,byts96-266-1} for physical applications)
was introduced by a number of authors as a convenient
 tool in order to deal with the evaluation of functional 
determinants.  It permits to give a meaning
in the sense of analytic continuation ---a mathematically very precise 
procedure--- to quantities that are formally divergent.

In this paper, we would like to revisit the zeta function regularization 
approach for the evaluation of expectation values $<O>$  and their quantum 
fluctuations. It is our 
opinion that zeta function regularization is a very powerful tool at our 
disposal, as compared with other methods and  that this issue deserves 
a careful investigation. 

We will mainly 
consider  two quantities: $O=\phi^2$ and the stress tensor trace 
$ O=T_\mu^\mu$
and their corresponding variances $\Delta^2O=<O^2>-<O>^2$. Within our 
formalism, it is convenient to introduce  the relative variance \cite{hu} 
\beq
\Delta_r^2O=\frac{\Delta^2 O}{<O>^2}\,.
\label{huo1}
\eeq
A different relative variance, though directly related to the previous one,
 has been introduced by Kuo and Ford \cite{ford1}. It reads
\beq
\tilde\Delta_r^2O=\frac{\Delta^2 O}{<O^2>}=\frac{\Delta_r^2O}{1+\Delta_r^2O}\,.
\label{huo}
\eeq

Considering now the operator $O=T_\mu^\mu$,  we observe that
the trace of the Einstein  equations reads
\begin{eqnarray} 
G_\mu^ \mu+D\Lambda= 8\pi G<T_\mu^ \mu >\:.
\end{eqnarray} 
Thus, fluctuations of the stress tensor trace induce classical  fluctuations
of the scalar curvature. 
Furthermore, this trace fluctuation may be used to have an estimate 
of the fluctuations of the whole  stress tensor, since
for conformally invariant quantum fields in 
 homogeneous space-times, one has
\begin{eqnarray} 
<T_{\mu \nu} >=\frac{ g_{\mu \nu}}{D} <T_\alpha^\alpha>\:.
\end{eqnarray} 
The issue concerning  the validity of 
semiclassical gravity has been discussed in \cite{ford1,hu001}.  

We  also recall that the first variation of the effective action is 
related to 
the vacuum expectation value of physical quantities, while the second 
variation of the effective action is associated with the variance.
Within the zeta function regularization procedure, one has to 
deal with traces of 
complex powers of elliptic operators. The first variation
of the effective action  
is well defined by the use of zeta function regularization, while the second 
variation is intrinsically ill defined, unless one makes use of suitable 
variations with disjoint supports. 
In the coincidence limit, the physically interesting case, one has to make use
of a further regularization \cite{hu}.  

Our main result is the following: modulo regularization problems, the relative 
variance turns out to be $\Delta_r^2O=\frac{2}{N}$, 
(thus $\tilde\Delta_r^2O=\frac{2}{N+2}$), 
where $N$ is the number of scalar fields in some multiplet. This result 
seems to be general, that is  independent on the quantity one is 
dealing with, for example the stress tensor trace. Our results 
are compatible and should be compared with the ones recently obtained, 
for $N=1$,  regarding the 
vacuum energy density fluctuations via smeared fields and point separation
\cite{hu00}, which give 
 relative variances of the order of unity albeit dimensionally dependent. 

On the other hand, coming back to the case of
 $N$ neutral scalar non-interacting fields, we  recover a 
well known 
criterion for neglecting the quantum gravity fluctuations in the large $N$ 
limit \cite{to77}.

The content of the paper is as follows. In  Sect. II,  zeta function 
regularization and heat-kernel techniques are briefly summarized. In Sect.
 III, the 
first variation of the effective action computed is shown to be related to 
the vacuum expectation values of observables. In Sect. IV, the second 
variation is shown to be associated with the variance and the final 
result is 
presented with the help of a further analytic and {\it ad hoc} regularization.
In Sect. V, some examples are presented.
The paper ends with some concluding remarks and an Appendix, where the first 
and second variations of the trace of the complex power of an elliptic operator
are explicitly computed.

\section{Zeta function regularization of the effective action}

In this section, we will summarize some basic aspects of the
heat-kernel and zeta function regularization methods.
 For the sake of simplicity we shall here
restrict ourselves to scalar fields, but the method is also valid in 
more general situations. In the case of a neutral scalar
field,  the one-loop Euclidean partition function,  reads
\cite{hawk77-55-133}
\begin{eqnarray} 
\Gamma^{(1)}=-\ln Z=\frac{1}{2}\ln\det
\frac{A}{\mu^2}\:,
\end{eqnarray} 
where  $\mu^2$ is a renormalization 
parameter.

To begin with, recall the definition of the zeta function related to an 
elliptic operator $A$ 
\begin{eqnarray}
\Gamma(s)\zeta(s|A)=
\frac{1}{\Gamma(s)}\int_{0}^\infty  t^{s-1} \mbox{Tr}\, e^{-tA}\, dt \,,
\label{h3}
\end{eqnarray}
which is valid for large values of $\Re s$.

For small $t$, and for second order  elliptic differential operators, 
it will be  assumed that the heat-trace has the following  asymptotics
\begin{eqnarray}
\mbox{Tr}\, e^{-tA}= \sum_{r=0}^\infty K_rt^{\frac{r-D}{2}}\,,
\label{h2}
\end{eqnarray}
where $K_r$ are the integrated heat-kernel coefficients. In principle they 
may be computed (see, for example \cite{klaus}).
We will also assume the validity of the local heat-kernel expansion, namely
\begin{eqnarray}
 e^{-tA}(x)= \sum_{r=0}^\infty K_r(x)t^{\frac{r-D}{2}}\,.
\label{h2l}
\end{eqnarray}  
If the asymptotic expansions (\ref{h2}) and (\ref{h2l}) hold, 
a standard argument leads to
the following meromorphic extension of the zeta function and its local 
counterpart ($J(s)$ being an analityc function)
\begin{eqnarray}
\Gamma(s)\zeta(s|A)=\sum_{r=0}^\infty \frac{K_r}{s+\frac{r-D}{2}}+J(s) \,.
\label{h4}
\end{eqnarray}
\begin{eqnarray}
\Gamma(s)\zeta(s|A)(x)=\sum_{r=0}^\infty \frac{K_r(x)}{s+\frac{r-D}{2}}
+J(s,x) \,.
\label{h44}
\end{eqnarray}

As a result, at $s=0$, the  global and local zeta functions are  regular and 
their derivatives exist.
We also have
$\zeta(0|A)(x)=K_{D}(x)$. As is well known,  $K_{D}(x)=0$ in any
odd dimensional manifold without boundary.
Under hypothesis above, 
zeta function regularization  corresponds to doing the following:
\begin{eqnarray} \ln \det 
\frac{A}{\mu^2}=-\zeta'(0|\frac{A}{\mu^2})=-\lim_{s \rightarrow 0}
\frac{d}{ds} \mbox{Tr}(\mu^{2s}A^{-s}) 
\:.\label{s3}
\end{eqnarray}

However we would like just to recall that, in some situations, 
it might be necessary to generalize the above definition in the form 
\begin{eqnarray} \ln \det 
\frac{A}{\mu^2}=-\lim_{s \rightarrow 0}\frac{1}{2}
\frac{d^2}{ds^2} \mbox{Tr}(s\mu^{2s}A^{-s}) 
\:.\label{s4}
\end{eqnarray}
When $\mbox{Tr}(A^{-s})$ is regular at $s=0$, the two definitions 
(\ref{s3}) and (\ref{s4}) coincide, but, in some cases, $\zeta(s|A)$
has a simple pole at $s=0$ and so Eq. (\ref{s3})
is no more valid.
This fact may well  be present at the 
level of  the effective action
(see for example \cite{cognola97,bytsenko01,elizalde01}). 

For the sake of simplicity, in the sequel we will assume
the validity of Eqs. (\ref{h4}) and (\ref{h44}). 
The latter gives  
\begin{eqnarray}
\zeta(0|A)(x)= K_{D}(x) \,.
\label{h4l0}
\end{eqnarray}

For future use, we also observe that when $K_{D-2}(x)\neq0$, 
the local zeta function has a simple pole at $s=1$. 
In fact we may write 
\begin{eqnarray}
\zeta(1+s|A)(x)=\frac{1}{\Gamma(1+s)}\:\frac{K_{D-2}(x)}s+G(1+s,x) \,,
\label{h4l1}
\end{eqnarray}
where $G(1,x)$ is a regular function. It is given by
\begin{eqnarray}
G(1,x)=\mbox{PP}\zeta(1|A)(x)-\gamma\:K_{D-2}(x) \,,
\label{h4l2}
\end{eqnarray}
$\gamma$ being the Euler's constant and 
\begin{eqnarray}
\mbox{PP}\zeta(1|A)(x)=\lim_{s \rightarrow 0}\left(\zeta(1+s|A)(x)- 
\frac{K_{D-2}(x)}{s}\right) \,,
\label{h4l89}
\end{eqnarray}
the finite part of $\zeta(s|A)$ at $s=1$.
When we consider an odd dimensional manifold without a boundary, 
this singularity is absent.

\section{The first variation of the effective action: 
vacuum expectation values}

In this section, we will evaluate vacuum expectation values $<O>$ of some 
specific quantities $O$, such as the stress tensor trace or conformal 
anomaly and the square of 
the field (fluctuation). These quantities involve the product of two quantum 
fields at the same point, $x$, and are therefore ill defined. They require a 
regularization.
We shall consider a multiplet of $N$ scalar fields denoted by $\phi$ 
in an external field, 
described through a classical action given by
\begin{eqnarray} 
S=\frac{1}{2}\int dx\:\phi L \phi \:,
\label{d0}
\end{eqnarray} 
where $L$ is a suitable (matrix valued) differential operator defined on a 
$D$-dimensional smooth manifold.  

To begin with, let us recall the formal trick that allows one to get the 
vacuum 
expectation value of the  bilinear $O=\phi K \phi$ within a path integral 
approach \cite{guido88}. 
We will consider two cases: $K=I$,the identity matrix, 
in the case of the field 
fluctuation $O=\phi^2$, and
$K=c_1+c_2 L$, in the case of stress tensor trace $O=T^\mu_\mu$. 
Here, $c_1$ and $c_2$ are 
constants and moreover,  in the conformally invariant case,
$c_1=0$. 
If we denote by $\alpha(x)$ a suitable classical source, we 
may consider the Euclidean partition function 
\begin{eqnarray} 
Z(\alpha) =\int D\phi e^{-\frac{1}{2}\int dx\:\phi L(\alpha)\phi} \:
=\left(\det \frac{L(\alpha)}{\mu^2}\right)^{-N/2}\,.
\label{d1}
\end{eqnarray} 
Here we are assuming that, in the massive case,  
the multiplet has the same common mass. In this way, there is no 
multiplicative anomaly (see, for example, \cite{eli3}) and
$L(\alpha)=L+\alpha K$ may be regarded as a simple differential operator.

A formal functional derivation  leads to 
\begin{eqnarray} 
<O(x)>=-2 \left. \frac{\delta \ln Z(\alpha)}{\delta  \alpha}
\right|_{\alpha=0} \:.
\label{d2}
\end{eqnarray} 
Making use of (\ref{d1}) and  zeta function regularization
(see Eq. (\ref{a5}) in the Appendix), 
we may give a meaning to the above formal
expression, namely
\begin{eqnarray}
\delta \ln Z(\alpha)=-\frac{N}{2}\lim_{s \rightarrow 0}\frac{d}{ds}
\left[\mu^{2s}s \mbox{Tr}\left(L^{-s-1}(\alpha)\delta L(\alpha) 
\right)\right]\,.
\label{d3}
\end{eqnarray}

In the case $O=\phi^2$, $\delta L(\alpha)=\delta \alpha$. Thus
\begin{eqnarray}
<\phi^2(x)>=N\lim_{s \rightarrow 0}\frac{d}{ds}
\left[s\mu^{2s} \zeta(s+1|L)(x)\right]\,.
\label{d4}
\end{eqnarray}
In the equations above, $\zeta(s|L)(x)$ is the local zeta function. 
As a result, making use of the meromorphic expansion (\ref{h4l1}), 
one gets (see \cite{moretti,moretti1,binosi})
\begin{eqnarray}
<\phi^2(x)>=N\mbox{PP} \zeta(1|L)(x)+NK_{D-2}(x)\ln \mu^2\,.
\label{d7}
\end{eqnarray}
When $D$ is odd and the manifold is without boundary, we simply have
\begin{eqnarray}
<\phi^2(x)>=N \zeta(1|L)(x)\,.
\label{d77}
\end{eqnarray}

In the other case, namely when one is dealing with the stress tensor trace, 
 $\delta L(\alpha)=\delta \alpha(c_1+c_2 L)$. 
As a consequence,
\begin{eqnarray}
<T^\mu_\mu(x)>=N\lim_{s \rightarrow 0}\frac{d}{ds}
\left[s\mu^{2s}(c_1 \zeta(s+1|L)(x)+c_2 \zeta(s|L)(x)\right]\,,
\label{d5}
\end{eqnarray}
and, as a result,
\begin{eqnarray}
<T^\mu_\mu (x)>=c_1<\phi^2(x)> +c_2N\,K_{D}(x)\,.
\label{d8}
\end{eqnarray}
In the conformally coupled case $c_1=0$ and one has the usual conformal 
anomaly, due only to  quantum effects.

\section{The second variation of the effective action: the variance}

We have seen that the first variation of the effective action is associated 
with  the vacuum expectation value  $<O>$ of bilinear quantities  in 
 quantum fields. 
Let us show now that the second variation of the effective action is related 
to the variance  $\Delta^2O=<O^2>-<O>^2$.

To begin with, the second variation of the partition function (\ref{d1}) gives 
\begin{eqnarray} 
<O(x) O(y)>= \left. 4\frac{1}{ Z(\alpha)} 
\frac{\delta^2 Z(\alpha)}{\delta \alpha(x)\delta \alpha(y) }
\right|_{\alpha=0} \:.
\label{d23}
\end{eqnarray} 
On the other hand, we have the identity 
\begin{eqnarray} 
\frac{\delta^2 \ln Z(\alpha)}{\delta \alpha(x)\delta \alpha(y) }=
\frac{1}{ Z(\alpha)} 
\frac{\delta^2 Z(\alpha)}{\delta \alpha(x)\delta \alpha(y) }-
\frac{1}{ Z^2(\alpha)} 
\frac{\delta Z(\alpha)}{\delta \alpha(x)}
\frac{\delta Z(\alpha)}{\delta \alpha(y)} 
\:.
\label{d24}
\end{eqnarray} 
As a consequence,  the variance is given by 
\begin{eqnarray} 
\Delta^2O(x,y)=<O(x) O(y)>-<O(x)><O(y)> =
4 \left.
\frac{\delta^2 \ln Z(\alpha)}{\delta \alpha(x)\delta \alpha(y) }
\right|_{\alpha=0}
 \:.
\label{d25}
\end{eqnarray} 
Now, it appears convenient to introduce the  relative variance
\begin{eqnarray} 
\Delta_r^2O(x,y)=\frac{\Delta^2O(x,y)}{<O(x)><O(y)>}
 \:.
\label{d26}
\end{eqnarray} 
The coincidence case, $x=y$, is particularly interesting from the 
physical point of view. It is formally given by
\begin{eqnarray} 
\Delta_r^2O(x)=\Delta_r^2O(x,y)|_{x=y}=
\frac{\delta^2 \ln Z(\alpha)}{\delta^2 \alpha(x)}               
\left(\frac{\delta \ln Z(\alpha)}{\delta \alpha(x)}\right)^{-2}
 \:.
\label{d267}
\end{eqnarray} 

Let us evaluate the second variation of $\ln Z(\alpha)$. In general, 
in the coincidence limit one gets  an 
ill defined quantity and a further regularization is required, as explained
in the Appendix.  
 Making use of (\ref{d3}), (\ref{s4}) and  (\ref{a6}) in the Appendix,
one has
\begin{eqnarray}
&&\delta_2\delta_1 \ln Z(\alpha)(\varepsilon)= 
\frac{N\mu^{2\varepsilon}}{2\Gamma^2(1+\varepsilon)}
\:\lim_{s \rightarrow 0}
\frac{d}{ds}
\left[\frac{\mu^{2s}}{\Gamma(s)}\:\times\right.\nn\\&&
\hspace*{-8mm}\left.
\int_0^\infty du\:u^{\varepsilon}
\int_0^\infty dv\:v^{\varepsilon} (u+v)^s 
\mbox{Tr}\left( e^{-uA}\delta_2 A e^{-vA}\delta_1 A \right)\right].
\label{2d3}
\end{eqnarray}
For $\varepsilon >0$ and sufficiently large, the integrand is regular at $s=0$,
and we have 
\begin{eqnarray}
\delta_2\delta_1 \ln Z(\alpha)(\varepsilon)&=&
\frac{N\mu^{2\varepsilon}}{2\Gamma^2(1+\varepsilon)}
\left[\int_0^\infty du\:u^{\varepsilon}
\int_0^\infty dv\:v^{\varepsilon}
\mbox{Tr}\left( e^{-uA}\delta_2 A e^{-vA}\delta_1 A \right)\right]\nonumber \\
&=&N\frac{\mu^{2\varepsilon}}{2}
\mbox{Tr}\left(A^{-1-\varepsilon}\delta_2 A A^{-1-\varepsilon}\delta_1 A 
\right)\,.
\label{2d33}
\end{eqnarray}
This is our general formula for the second variation of the
 functional determinant.
It is in agreement with a result obtained in \cite{kont95b}. 

Let us consider the specific variation related to our observables. 
First, for $O=\phi^2$, $\delta L(\alpha)=\delta \alpha$, and 
in the coincidence limit we have 
\begin{eqnarray} 
<\Delta^2\phi^2>(\varepsilon)
=2N\mu^{2\epsilon}\zeta^2(1+\varepsilon| L)(x).
\label{d27}
\end{eqnarray} 
For $x \neq y$, $\zeta(1+\varepsilon|L)(x,y)$ is regular at $\varepsilon=0$,
but in the
coincidence limit $y \rightarrow x$, $\zeta(1|L)(x)$ is well 
defined only for odd $D$ and boundary free manifolds.   

In the conformally coupled case and for $O=T_\mu^\mu$, one has  
  $\delta L(\alpha)=c_2\delta \alpha L$. As a result
\begin{eqnarray} 
<\Delta^2T^\mu_\mu>(\varepsilon)=
2Nc_2^2\zeta^2(\varepsilon| L)(x)
=2Nc_2^2\zeta^2(0|L)(x)+O(\varepsilon)
\end{eqnarray} 
and we can remove the regularization parameter 
because  $\zeta(s|L)(x)$ is  regular at the origin.

Some remarks are in order. 
If we limit ourselves to the odd dimensional case,
in a manifold without boundary, 
the analytic continuation works in a simple manner 
for both the quantity $O=T_\mu^\mu$ and $O=\phi^2$
and no scale dependence appears in the final expressions. As a 
consequence, it turns out that the relative variance,
\begin{eqnarray} 
\Delta_r^2O=\frac{<O^2>-<O>^2}{<O>^2}
\,,
\end{eqnarray} 
is always equal to
\begin{eqnarray} 
\Delta_r^2O=\frac{2}{N}\,,
\end{eqnarray} 
and
\begin{eqnarray} 
\tilde\Delta_r^2O=\frac{2}{N+2}
\,.
\end{eqnarray} 

The expressions above are valid for a multiplet of scalar $N$ fields.
In the case of other quantum fields, the formula for the relative
variance could be a little bit different 
(see for example ref.~\cite{kaza02} for the treatment of gauge fields).

In the even dimensional case, or/and in the presence of boundary, 
the situation is more complicated and 
a further renormalization seems unavoidable.

\section{Examples}
 
\subsection{$D-$dimensional torus}
 
In this Section we will consider, as first example, the $D-$dimensional torus.
This is a symmetric flat manifold with finite volume and the local zeta 
function is simply the ratio of the global zeta function and the torus volume. 
Thus, we may limit ourselves  to the discussion involving the zeta function.
This is a general conclusion valid for every symmetric space.

The zeta function for  the case of a massive ($N=1$) scalar field is given by
\beq
\zeta_D (s|L) = \left( \frac{4 \pi^2}{R^2} \right)^{-s} Z_D (s; (mR)^2)
\eeq
with \cite{eejpa01}--\cite{bek1}
\beq
Z_D (s; (mR)^2)& =& (2\pi )^{D/2} (mR)^{D-2s}
\frac{\Gamma(s-D/2)}{\Gamma (s)} + \frac{2^{s/2+D/4+1}\pi^s
(mR)^{-s +D/2}}{\Gamma (s)} \:\times\nn \\ && \hspace*{0mm}
 {\sum_{\vec{n} \in \mbox{\bf Z}^D}}' \left( \vec{n}^2 \right)^{s/2-D/4}
\, K_{D/2-s} \left( 2^{3/2}\pi mR \sqrt{ \vec{n}^2}\right). \label{tm1}
\eeq
Here $R$ is the radius of the torus and $m$ the mass of the field.
This expansion is exponentially convergent. It is to be seen not just as a
big mass expansion (the convergence is then extremely fast), for it
is valid in a very wide range of values of $mR$: $1\simeq mR < \infty$.
 
In the case of a massless field, the convenient expression to be used is
quite different (this is explained  in detail in \cite{eejpa01}):
\beq
Z_D (s)& =&  \left. \frac{2^{1+s}}{\Gamma (s)} \sum_{j=0}^{D-1}\right[
\pi^{j/2} \Gamma(s-j/2) \zeta_R (2s-j)   \nn\\
&&  \left.   + 4\pi^s \sum_{n=1}^\infty
{\sum_{\vec{n}_j \in \mbox{\bf Z}^j}}' n^{j/2-s}
\left( \vec{n}^2 \right)^{s/2-j/4}
\, K_{j/2-s} \left( 2\pi n \sqrt{ \vec{n}_j^2}\right)\right].\label{tm0}
\eeq

On the other hand, for the sake of completeness,
it is interesting to have a perturbative  expression for
the case when the mass of the field is very small but different from zero.
This is obtained by means of binomial expansion in the equation defining
the zeta function. The result is:
\beq
Z_D (s; (mR)^2) = \sum_{k=0}^\infty (-1)^k \frac{\Gamma (s+k)}{k! \,
\Gamma (s)} (mR)^{2k} Z_D (s+k). \label{tm10}
\eeq
This expression, combined with the preceding one, Eq. (\ref{tm0}), yields the
desired low mass expansion. Such expansions for the zeta function
are not to be found in the literature. In fact, explicit expressions of
the type (\ref{tm0}) have appeared in the seminal paper \cite{eejpa01}
for the first time. They are convenient, on the one hand, because they
exhibit the pole structure of the zeta function explicitly. On the other,
they are useful from the computational point of view, because
they  consist of a term including the main contribution, together
with a series that converges extremely fast: only a few first terms need
to be computed in order to obtain results as accurate as desired.
 
From these expressions, the first and second variations of the effective
action are immediate to compute. Essentially, what we get are expressions
of the following kind: (i) in the massless case ($p$ is a natural number, 
a small one for any of the operators here considered)
\beq
&& \frac{Z_D (p)}{[Z_D (p-1)]^2} = \frac{\Gamma (p-1)^2 }{2^{p-2}\Gamma (p)}
\:\times\\  && \hspace*{-3mm}
\frac{\sum_{j=0}^{D-1} \left[ \pi^{j/2} \Gamma(p-j/2)
\zeta_R (2p-j) + 4\pi^2 S_{D,j}(p) \right]}
{\left\{ \sum_{j=0}^{D-1} \left[ \pi^{j/2} \Gamma(p-1-j/2) \zeta_R (2p-2-j)+
4\pi^2 S_{D,j}(p-1) \right] \right\}^2}, \nn
\eeq
where the $S_{D,j}(p)$  are fast convergent series providing only
corrections to the main terms, (ii)
\beq
&& \frac{Z_D (p; (mR)^2)}{[Z_D (p-1; (mR)^2)]^2} =  \\  &&
\hspace*{-2mm} \frac{Z_D (p)}{[Z_D (p-1)]^2}
\left[ 1 + 2p \left(\frac{Z_D (p)}{[Z_D (p-1)]^2} -
 \frac{Z_D (p+1)}{[Z_D (p)]^2}\right)(mR)^2 + \cdots \right], \nn
\eeq
in the case of a field of very small mass (after doing a small-mass
expansion as described above), and (iii)
\beq 
&&\frac{Z_D (p; (mR)^2)}{[Z_D (p-1; (mR)^2)]^2} = 
\frac{(mR)^{2p-4}\Gamma (p-1)^2}{(2\pi )^{D/2} \Gamma (p)}
\:\times\\  && \hspace*{-10mm}
  \frac{\left[ \Gamma(p-D/2) + 2^{p/2-D/4+1}\pi^{p-D/2} (mR)^{p -D/2}
+ \Sigma_{D}(p)\right]}
{\left[\Gamma(p-1-D/2)
+ 2^{p/2-D/4+1/2}\pi^{p-1-D/2} (mR)^{p-1 -D/2} + \Sigma_{D}(p-1)\right]^2},\nn
\eeq 
in the massive case, where again $\Sigma_{D}(p)$ is a  very
fast convergent series. These expressions simplify very much when poles of the
gamma function appear (for even dimension $D$). They are quite easy to 
deal with. As advanced before, no singularity appears for $D$ odd, and the
whole expression remains, in that case (the series being well approximated by
a couple of terms).

\subsection{The Casimir slab}

As a second example, we will revisit the computation of the local zeta function
related to the  Casimir slab, namely a massless scalar 
quantum field confined in the $x$ direction between infinite parallel planar
Dirichlet boundaries, located at $x=0$ and $x=a$. In this case, 
 the local zeta function 
is not trivial, due to the presence of the planar boundaries 
(see, for example \cite{moste,fulling,eliz94b,eli2}).   

To start with, recall the form of the local diagonal heat-kernel, which 
depends only on 
the confining coordinate $x$, i.e.
\beq
<x|e^{-tL}|x>=\frac{2}{a(4\pi t)^{\frac{D-1}{2}}}\sum_{n=1}^\infty
\sin^2 \frac{n\pi x }{a}\:\exp{-\frac{n^2\pi^2\:t}{a^2})}\,.
\eeq
Here $L=-\nabla^2$ is the Dirichlet Laplacian in the slab. Making use of the
trigonometric duplication formula for the sine and 
Poisson-Jacobi re-summation formula, we may rewrite the above expression in the
form  
\beq
<x|e^{-tL}|x>=\frac{1}{(4\pi t)^{\frac{D}{2}}}
\left( \sum_{n=-\infty}^{\infty}
\exp{(-\frac{a^2 n^2}{t})}-
 \sum_{n =-\infty}^\infty \exp{(-\frac{(na+x)^2}{t})}\right)
\,.
\eeq
All the Seeley-DeWitt $K_r(x)$ coefficients  
vanish, but the first one $K_0(x)=1$. As a consequence, we may 
anticipate that $<T_\mu^\mu(x)>$ is vanishing. 
Making use of the  Mellin transform and the above expression for the 
heat-kernel,  and analytically regularizing the integral \cite{gelf}
\beq
\int_0^\infty dt \ t^z=0\,,
\eeq
the analytic continuation for the local zeta 
function may be obtained. We present the result in a simple and 
symmetric form (for another equivalent form corresponding to $D=4$, 
see \cite{moretti})
\beq
&& \zeta(s|L)(x)= 
\frac{\Gamma(\frac{D}{2}-s)a^{2s-D}}{(4\pi)^{\frac{D}{2}}\Gamma(s)}
\:\times\label{guido}\\ && 
\:\:\left[ 2\zeta_R(D-2s)-\zeta_H(D-2s,x/a)-
\zeta_H(D-2s,1-x/a)\right]\,. \nn 
\eeq
In the above expression, $\zeta_R(z)$ and $\zeta_H(z,q)$ are the Riemann and
Hurwitz zeta functions respectively. This result can also be obtained by 
making use of the re-summation techniques explained in \cite{actor,eliz94b}.  

Some comments are in order. The local expression we have obtained is 
already in the renormalized form.  
Furthermore, the related trace involves 
a (infinite, but trivial) volume $V_T$ in the transverse directions and an 
integration over $x$. Since
\beq
\int_0^1 dq \ \zeta_H(z,q)=0\,,
\eeq 
the resulting zeta function reads
\beq
\zeta(s|L)=
\frac{2V_T\:\Gamma(\frac{D}{2}-s)a^{2s-D+1}}{(4\pi)^{\frac{D}{2}}\Gamma(s)}
\zeta_R(D-2s)\,,
\label{guido1}
\eeq 
and this is exactly the zeta function associated with the Casimir slab 
configuration. 

As far as the application to the evaluation of the vacuum expectation value 
is concerned, first, let us consider  $D >2$. The zeta function is regular at
$s=1$ and the result is  
\beq
&& <\phi^2(x)>= \frac{\Gamma(\frac{D}{2}-1)a^{2-D}}{(4\pi)^{\frac{D}{2}}}
\:\times\label{guido3} \\ &&
\:\:
\left[ 2\zeta_R(D-2)-\zeta_H(D-2,x/a)-
\zeta_H(D-2,1-x/a)\right]. \nn
\eeq
The expression is finite everywhere and gives a 
vanishing result at $x=0$ and $x=a$. Due to the simple geometry of the planar 
boundaries, the boundary divergences are not present.  

Things are different in $D=2$. In this case, the zeta function has a pole  
at $s=1$ and 
the dependence on the scale $\mu$ appears. For the sake of 
completeness, we give the result. It reads
  
\beq
<\phi^2(x)>=
\frac{1}{2\pi}
\left[\gamma+\ln \frac{a \mu}{\pi}+\ln\sin\frac{\pi x}{a}\right]\,.
\label{guido4}
\eeq
In this case, the boundary divergences are present.

A few other exact analytic results obtained via zeta function regularization 
can be found in \cite{moretti}. 

\section{Concluding remarks}

In this paper, we have revisited the use of zeta function regularization 
approach to the evaluation of expectation values of physical quantities  and 
their related quantum variances. The former are associated with 
the evaluation of the first variation of the effective action, while the 
latter are related to the second variation. We have shown that the first 
variation can be regularized by use of zeta function techniques, and 
explicit  expressions for the vacuum expectation values have been 
exhibited.  For issues concerning the second variation,
in general, zeta function regularization works well only 
when one is dealing with 
off diagonal terms, since the coincident limit is highly singular. A 
further analytic regularization has to be introduced to treat the coincidence 
limit. 

We may summarize our results as follows. For a multiplet of $N$ scalar 
fields and $O=T_\mu^\mu$,  in the conformally 
coupled case,  analytic regularization gives a relative variance exactly 
equal to  $\frac{2}{N}$, independently on the dimension of the boundary free 
manifold. 
For $O=\phi^2$, we have obtained again a relative variance exactly equal to 
$\frac{2}{N}$, but now limited to the odd dimensional case without boundary. 
These restrictions can be removed in some particular situations, like 
the case of the Casimir slab, where one is 
dealing with a flat manifold with flat boundaries. The even dimensional case
seems to require however further renormalization.      

The formalism can be directly applied to the
expectation value of the stress tensor (see \cite{hu}), but only after much 
more effort and work. Again, the first variation of the effective action with 
respect to the metric tensor gives $<T_{\mu \nu}> $. With regard to this issue,
for an evaluation using local zeta function regularization see 
\cite{moretti1,moretti2}.

The second variation is related to the variance, and for the off diagonal 
case, there
is again factorization. In the special case of homogeneous space-times 
and for a  conformally coupled scalar field, one recovers the $O=T_\mu^\mu$ 
case.    

\section{Appendix}

In order to compute the variation of the trace of an elliptic invertible 
operator $A$,  one has to  take into 
account the fact that the variation (deformation) of $A$ does not commute 
with $A$.
From  $ A^{-1}A=I$,  we have
\begin{eqnarray}
\delta A^{-1}=-A^{-1}\delta A A^{-1}\,.
\label{a1}
\end{eqnarray}
For the calculation of complex powers of $A$ we can use the Cauchy-Dunford 
representation 
\begin{eqnarray}
A^{-s}=-\frac{1}{2\pi i}\int_C dz z^{-s}(A-z)^{-1}\,,
\label{a2}
\end{eqnarray}  
in which $C$ is a suitable contour on the  complex $z-$plane.
As a consequence, we have
\begin{eqnarray}
\delta A^{-s}=\frac{1}{2\pi i}\int_C dz z^{-s}(A-z)^{-1}\delta A  (A-z)^{-1}\,.
\label{a3}
\end{eqnarray}  
Making use of the two representations (for $\mbox{Re } z >0$  and 
$\mbox{Re } s >0$):
\begin{eqnarray}
\Gamma(s)(A-z)^{-s}=\int_0^\infty dt t^{s-1}  e^{tz} e^{-tA}\,,
\label{a22}
\end{eqnarray}  
and 
\begin{eqnarray}
\frac{1}{\Gamma(z)}=-\frac{1}{2\pi i}\int_C dw (-w)^{-z} e^{-w}\,,
\label{a23}
\end{eqnarray}  
one gets
\begin{eqnarray}
\delta \mbox{Tr}A^{-s}=-s \mbox{Tr}(A^{-s-1}\delta A) \,.
\label{a5}
\end{eqnarray}  
If the operator $A$ is self-adjoint, then there exist eigenvalues and  
eigenvectors,  $\lambda_n$ and $\Psi_n$, such that 
$A \Psi_n=\lambda_n\Psi_n$, and we have   
\begin{eqnarray}
\delta \mbox{Tr}A^{-s}&=&\frac{1}{2\pi i}\int_C dz z^{-s} 
\mbox{Tr}((A-z)^{-2}\delta A) \nonumber \\
&=& \frac{1}{2\pi i} \int_C dz z^{-s} \sum_n \frac{1}{(z-\lambda_n)^2}
<\Psi_n,\delta A \Psi_n> \nonumber \\
&=&
-s\sum_n \lambda_n^{-s-1}<\Psi_n,\delta A \Psi_n>
=-s \mbox{Tr}(A^{-s-1}\delta A) \,.
\label{a55}
\end{eqnarray}

For the second variation, making use again of (\ref{a22}) and
(\ref{a23}), one gets 
\begin{eqnarray}
\delta_2 \delta_1 \mbox{Tr}A^{-s}&=&-s \mbox{Tr}(\delta_2A^{-s-1}\delta_1 A)
\nonumber \\
&=&-s\frac{1}{2\pi i}\int_C dz z^{-s-1} 
\mbox{Tr}((A-z)^{-1}\delta_2 A (A-z)^{-1}\delta_1 A)\,.
\label{a66}
\end{eqnarray}  
This expression is valid only if $\mbox{Supp}\,\delta_1$ has void intersection 
with  
$\mbox{Supp}\,\delta_2$. Since we are interested in the coincidence limit 
$y=x$, we have to introduce an additional
 analytic regularization. The simplest one 
is to replace $(A-z)^{-1}$ with $ (A-z)^{-1-\varepsilon}  $ and employ 
the representation in terms of the Mellin transform:
\begin{eqnarray}
\Gamma(1+\varepsilon)(A-z)^{-1-\varepsilon}=
\int_0^\infty du\, u^{\varepsilon}\,  e^{uz} e^{-uA}\,,
\label{a22c}
\end{eqnarray}  
in Eq. (\ref{a6})  and then use again (\ref{a23}). As a result, 
we arrive to the formula 
\begin{eqnarray}
\hspace*{-3mm} \delta_2 \delta_1 \mbox{Tr}A^{-s}(\varepsilon)=
-\frac{1}{\Gamma^2(1+\varepsilon)}
\frac{1}{\Gamma(s)}\int_0^\infty du u^{\varepsilon}\int_0^\infty dv
v^{\varepsilon} (u+v)^s 
\mbox{Tr}\left( e^{-uA}\delta_2 A e^{-vA}\delta_1 A \right)\,.
\label{a6}
\end{eqnarray}

This representation for the second variation of the trace of a complex power
of an elliptic operator $A$ is in agreement with the one obtained in 
\cite{hu}, making use of a different method based on Schwinger's perturbative
expansion.

\section{Acknowledgments}
We are grateful to Valter Moretti for a useful discussion.
This investigation have been supported by the cooperation agreement
INFN (Italy)--DGICYT (Spain). EE has been financed also by DGI/SGPI (Spain), 
project BFM2000-0810, and by CIRIT (Generalitat de Catalunya), contract 
1999SGR-00257.

\end{document}